\documentclass{ptapap}

\author{{\L}ukasz Wyrzykowski}[OAUW]
\author{Zuzanna Kostrzewa-Rutkowska}[OAUW]
\author{Krzysztof Rybicki}[OAUW]
\affil[OAUW]{Warsaw University Astronomical Observatory, Al. Ujazdowskie 4, 00-478 Warszawa, Poland}

\title{Microlensing by single black-holes in the Galaxy}

\begin{document}

\maketitle

\begin{abstract}

The longest microlensing events provide enough information to estimate the mass and distance of the lens. Among hundreds of millions of stars which were monitored for many years by the OGLE project we selected those with clear parallax effect and derived the mass function of lensing objects in the Milky Way. We also found candidates for microlensing stellar-mass single black holes. We discuss how Gaia superb astrometry will help in measuring masses of remnants in currently on-going and future microlensing events.

\end{abstract}

\section{Introduction}

Currently, there is about 30  stellar-mass black holes candidates known in the Milky Way.
All of them, however, were discovered in binary systems with another star thanks to the X-ray emission of the accretion process. 
The black holes detected via this channel have masses above 6 Solar masses, as their mass function has a clear cut off \citep{Ozel2010}. 
On the other hand, other end-products of stellar evolution for slightly less massive progenitor stars, neutron stars, do not exceed 2 Solar masses and have a narrow mass distribution \citep{Kiziltan2013}. 
Therefore, observations indicate a clear gap in the mass distribution of stellar remnants, which is, on the other hand, not reproduced by theoretical predictions and simulations of stellar evolution (eg. \citealt{FryerKalogera2001}). 
The synthetical mass distribution shows a peak for neutron stars and then gradually decline with increasing masses of black holes. The discrepancy with observations is a sign that either there is something wrong in our understanding of the basics of the stellar evolution, or the observational biases lead to an artificial gap in mass distribution.
Detection of isolated black holes and a complete census of masses of stellar remnants is therefore crucial for a robust verification of theoretical predictions of stellar evolution. 

Gravitational microlensing phenomenon \citep{Paczynski1996} gives a unique opportunity for solving this intriguing astrophysical problem since it is sensitive solely to the mass of the lensing object and it does not require any light to be emitted by the lens.
Predictions by \cite{Gould2000} show that about 20\% of all microlensing events should be due to stellar remnants, with 17\% caused by white dwarfs, 3\% by neutron stars and 0.8\% should have black hole lenses. 
Within more than 2000 microlensing events found every year by the OGLE project there should be about 16 black holes causing lensing.
However, the main difficulty in recognising the nature of the lens is because of strong degeneracy in mass and distance computation for a standard microlensing event. 
The mass of the lens can be computed if apart from the standard time-scale of the event ($t_E$) is measured, but also the information on relative velocity or proper motion between the lens and the source as well as lens and source distance is known. 
For a standard microlensing event solely the time-scale is determined, hence the mass is not measurable for an individual event.

However, if there is an additional effect detectable in the light curve, namely the microlensing parallax, and we are able to somehow measure the angular Einstein radius of the lensing configuration, the mass can be computed with no ambiguity. 
So far, however, such lucky configurations that we had both informations happened in case of very few microlensing events.
In order to find candidates for heavy lenses we searched for microlensing parallax signal in microlensing events found among hundreds of millions of stars monitored by the Polish Optical Gravitational Lensing Experiment (OGLE). 
This work briefly reminds the results from \cite{WyrzykowskiBH}.

\section{Search}
The third phase of the OGLE project (OGLE-III) operated from 2001 until 2009 and used the 1.3m Warsaw Telescope, located at the Las Campanas Observatory, Chile, operated by the Carnegie Institution for Science.
OGLE-III observed the Galactic Bulge in 177 fields; 0.34 sq.~deg. each over 8 mosaic CCDs. 
We selected the best observed 91 fields covering, in total, 31 sq.~deg. and containing about 150 million sources observed in the I-band down to 20.5 mag.
The search procedure included computation of simple statistics on light curves as well as fitting the microlensing curve model without and with parallax effect. 
The Earth-based microlensing parallax is caused by the motion of the observer on Earth during the duration of the event (hence changing the observed amplification). This causes the light curve to deviate from the standard bell-shaped and symmetrical Paczynski curve. 
The deviating light curves were selected using Random Forest Machine Learning algorithm trained on a visually prepared sample of parallax and standard events from our previous search for standard events \citep{Wyrzykowski2015}.

\begin{figure}[h]
\centering
\includegraphics[width=7cm]{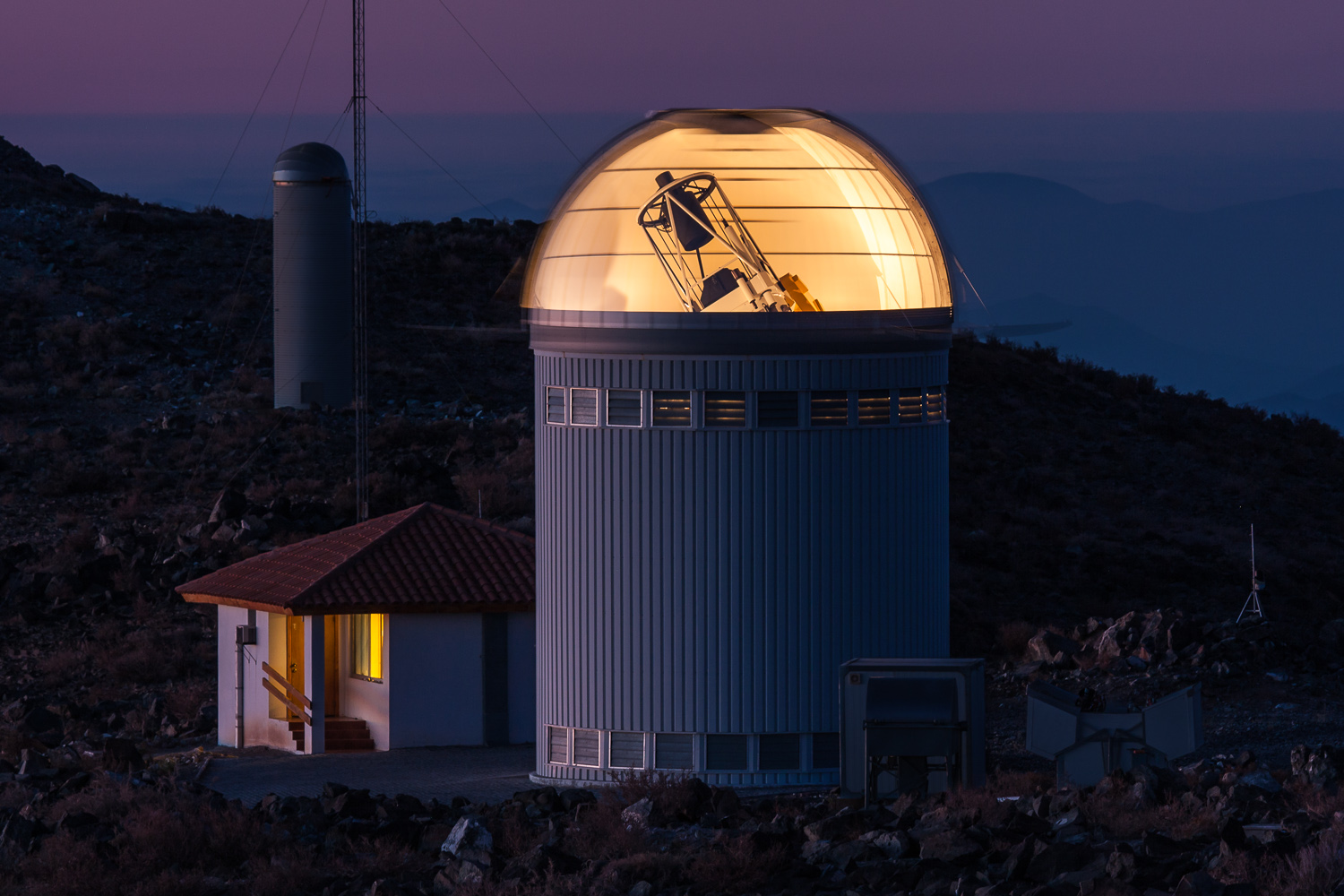}
\caption{OGLE telescope in Las Campanas, Chile, in operation since 1996. Photo by K.Ulaczyk and L.Wyrzykowski.}
\label{fig:ogletel}
\end{figure}

\section{Candidates for dark remnants}
We have automatically selected 59 high quality parallax events. Those were modelled using MCMC in order to recognise all possible degenerate solutions for microlensing parameters. 
Next, based on the available colour (V-I) information, we selected 26 events, which had sources located in the Galactic Bulge, i.e., were located in the Red Clump Region on the colour-magnitude diagram for sources. This allowed for constraining the distance to the source at $\sim$8 kpc as well as allowed us to use proper motion distribution for the sources typical for the Galactic Bulge. 
For the lens we assumed proper motion distribution for the Galactic Disk; all proper motion distributions were taken from HST studies from \cite{Calamida2014}.

By supplementing microlensing parameters obtained from photometry with priors on missing parameters, we obtained the posterior probability distributions for mass and distance for 26 lenses. 
We then computed the expected brightness of the lens if it was a main sequence star for a given mass and distance, and compared them to the observed blended light seen in the microlensing curve. 
In other words, if the lens was luminous then we should be able to accommodate its light into the blended light, typically composed of the source, lens and additional light from stars not related with the event. Based on the comparison and the probability density we computed the probability for an event to have a dark lens. There were 19 events with probability larger than 50\%, 15 events with probability larger than 75\% and 3 events with probability larger than 95\%. 

\begin{figure}[h]
\includegraphics[width=\textwidth]{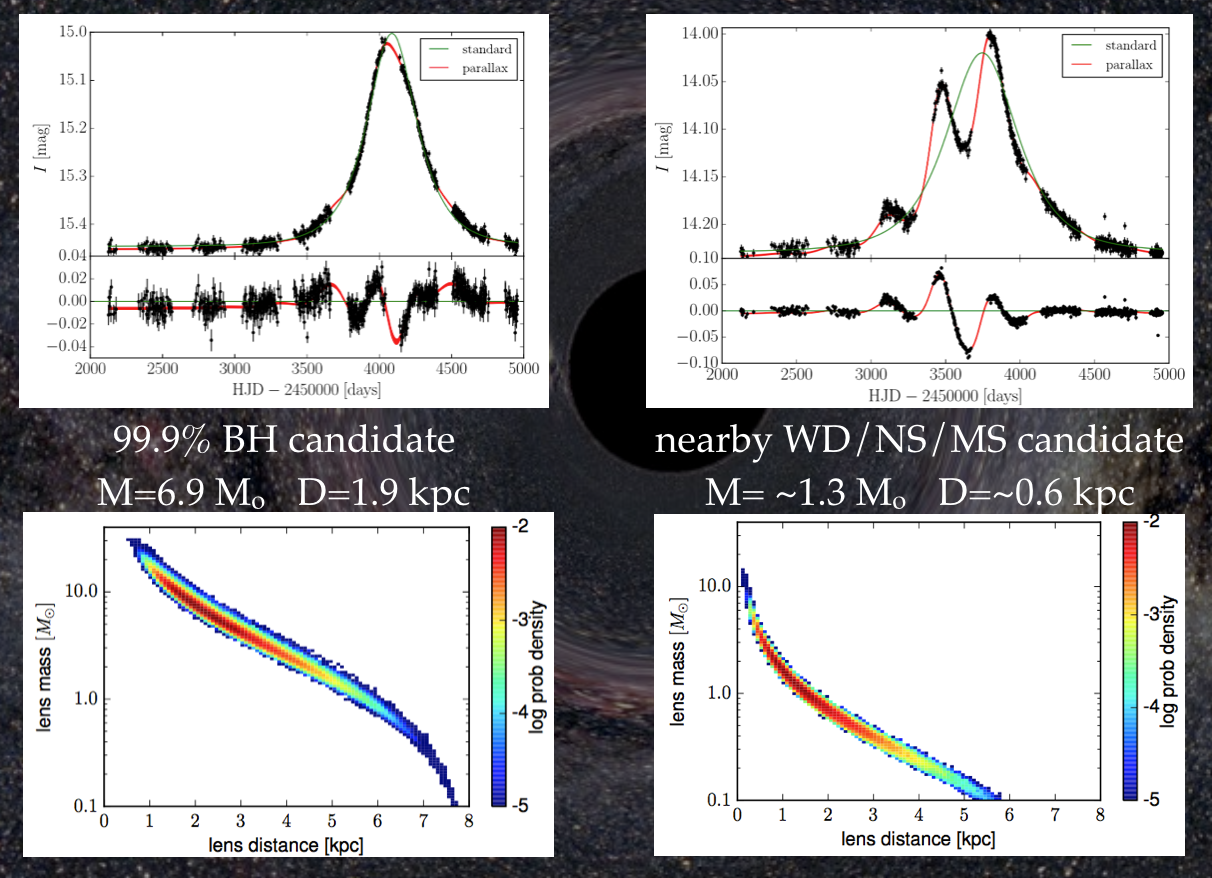}
\caption{OGLE3-ULENS-PAR-02 (left panels) and OGLE3-ULENS-PAR-01 (right panels) microlensing events with strong parallax signal. PAR-02 is our best candidate for a black hole lens with the most probable mass of 6.9 solar masses. PAR-01 is potentially an interesting nearby white dwarf candidate. Top panels show light curve with the parallax model and its residuals. Bottom panels show 2D probability distributions for mass and distance of the lens.}
\label{fig:odravistula}
\end{figure}

Figure 2 shows two events with the strongest parallax signal for which the mass and distance were best constrained. 
PAR-02 is our best candidate for a black hole lens, as the mass probability density reaches its peak at around 7 solar masses (lower left plot on Fig.2) and therefore the chances the lens is luminous at such mass and distance is very low. 

\section{Gap or Not Gap}
Figure 3 shows probability distributions for masses of our dark lens candidates. Since they are based on assumptions on unknown relative proper motion, they are relatively wide. However, the distribution of their medians (thin blue lines and histogram in the bottom panel) hint there is a continuous distribution of masses of remnant lenses in the range between 1 and 10 solar masses.
This indicates lack of, so-called, {\it mass gap} in the mass distribution of neutron stars and black holes, present in X-ray binaries.

\begin{figure}[ht]
\centering
\includegraphics[width=9cm]{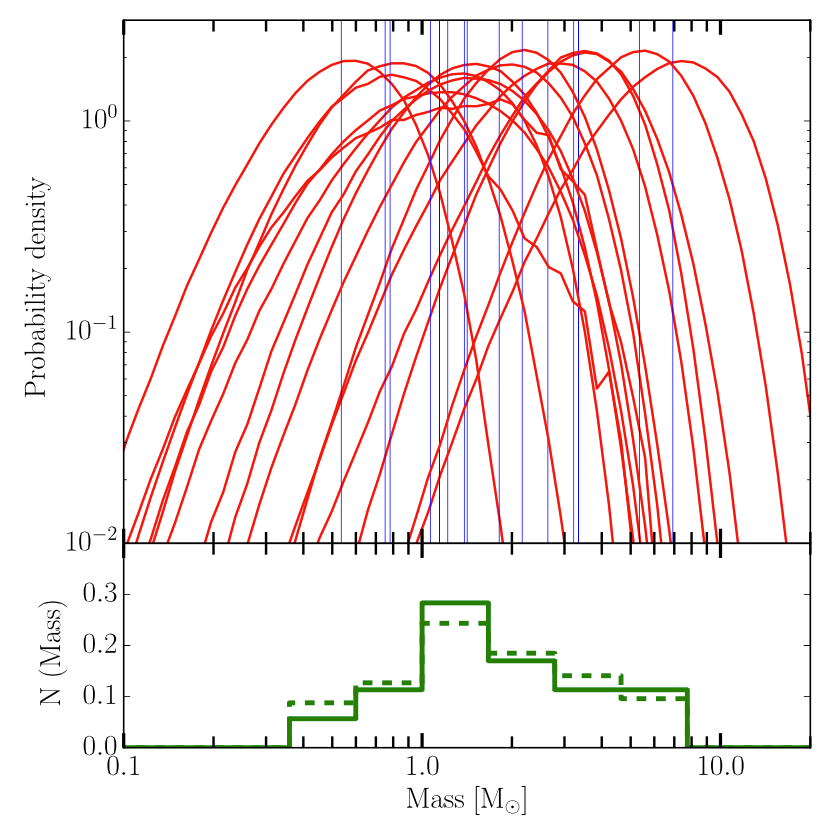}
\caption{Probability distributions of masses for 15 best candidates for dark lenses. Blue lines indicate medians. The bottom panel shows histogram for medians with and without efficiency correction (dotted vs. solid line, respectively). The distribution of masses between 1 and 10 seems to be continuous, hinting lack of mass gap in heavy remnants distribution.}
\label{fig:massfunc}
\end{figure}

\section{Astrometric microlensing with Gaia}
Photometry alone, even with parallax effect, can not provide unambiguous mass determination in a microlensing event and additional astrometric data are needed.
Several searches for those minute astrometric centroid displacements are being conducted, however, so far none were reported positive. 
In the near future, however, the Gaia space mission (e.g., \citealt{deBruijne2014}) will provide sub-milliarcsecond positional time-series for a billion stars in our Galaxy, down to $V\sim20$ mag. 
Such precise astrometry will be most suitable for detecting and measuring microlensing astrometric signals (e.g., \citealt{Wyrzykowski2012}). 
Full Gaia mission data with its astrometry will be available in a few years and will be sufficient to determine mass for numerous ground-based found events with measured microlensing parallax. 
Our simulations of expected astrometric signal for already on-going microlensing events found in OGLE show that combination of OGLE photometry and ground-based parallax measurement with Gaia astrometry will yield mass determination of lenses with accuracy better than 20\% (Rybicki et al. in prep.).

\begin{figure}[ht]
\centering
\includegraphics[width=9cm]{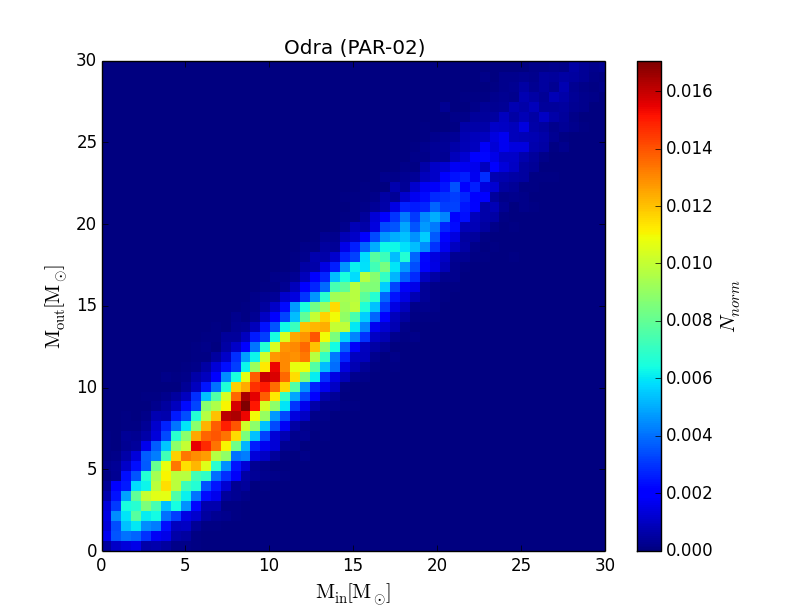}
\caption{Example result of simulations of astrometric observations of Gaia combined with OGLE ground-based photometry for OGLE3-ULENS-PAR-02 dark lens (black hole) candidate. Mass of the lens can be determined better than 20\%.}
\label{fig:astrom}
\end{figure}

\section{Conclusions}
We searched a vast data base of OGLE-III photometric light curves of 150 million stars towards the Galactic bulge and found 3 strong candidates for microlensing events caused by stellar-mass black holes located in the Galactic disk. The distribution of masses of other dark lenses indicate lack of mass gap between neutron stars and black holes. The black hole candidates and masses of lenses still need to be confirmed with additional ground- and space-based follow-up observations (X-rays, adaptive optics astrometry). In near future Gaia will provide accurate astrometric measurements of microlensing events and thus measuring masses of dark lenses will be possible with much better precision.

\acknowledgements{We acknowledge the contributions of the entire OGLE group to this study. We also thank Gaia DPAC, in particular Jos de Bruijne.
This work is supported by the Polish NCN 'Harmonia' grant No. 2012/06/M/ST9/00172.}

\bibliographystyle{ptapap}
\bibliography{wyrzykowski-bh}

\end{document}